\documentclass[useAMS,usenatbib,fleqn]{mnras}
\usepackage{calrsfs}
\DeclareMathAlphabet{\pazocal}{OMS}{zplm}{m}{n}
\usepackage{multirow}
\usepackage{hhline}
\usepackage{times}
\usepackage{graphics,epsf}
\usepackage[T1]{fontenc}
\usepackage{aecompl} 
\usepackage{amsmath}               
\usepackage{placeins}
\usepackage{booktabs}
\usepackage{amsfonts}               
\usepackage{amssymb}                
\usepackage{epsfig}                 
\usepackage{epstopdf}
\usepackage{rotating}
\usepackage{color}
\usepackage{tabularx}
\usepackage{url}

\def \apj{ApJ}
\def \aap{A\&A}
\def \aj{AJ}

\def \mnras{MNRAS}
\def \apjl{ApJ Lett.}
\def \apjs{ApJ Suppl. Ser.}
\def \nat{Nature}

\def \icarus{Icarus}

\begin{document}

\title[Capture of interstellar objects] {Planet seeding through gas-assisted capture of interstellar objects }

\author[Grishin et. al.]{
Evgeni Grishin,$^{1}$
Hagai B. Perets,$^{1}$
Yael Avni $^{1}$
\\
$^{1}$Physics Department, Technion - Israel institute of Technology, Haifa,
Israel 3200002\\
E-mail: eugeneg@campus.technion.ac.il (EG);  hperets@physics.technion.ac.il (HBP);
yael.avni10@gmail.com (YA)}
\maketitle

\begin{abstract}
Planet formation begins with collisional growth of small planetesimals accumulating into larger ones. Such growth occurs while planetesimals are embedded in a gaseous protoplanetary disc. However, small-planetesimals experience collisions and gas-drag that lead to their destruction on short timescales, not allowing, or requiring fine tuned conditions for the efficient growth of $\sim$ metre-size objects. Here we show that $\sim10^{4}$ interstellar objects such as the recently detected 1I/2017-U1 ('Oumuamua) could have been captured, and become part of the young Solar System, together with up to hundreds of $\sim$ km sized ones. The capture rates are robust even for conservative assumptions on the protoplanetary disc structure, local stellar environment and planetesimal ISM density. 'Seeding' of such planetesimals then catalyze further planetary growth into planetary embryos, and potentially alleviate the main-challenges with the meter-size growth-"barrier". The capture model is in synergy with the current leading planet formation theories, providing the missing link to the first planetesimals.  Moreover, planetesimal capture provides a far more efficient route for lithopanspermia than previously thought. 
\end{abstract}


\section{Introduction}

The early stages of planet formation are thought to occur in gaseous protoplanetary-discs (PPD). The primordial PPD consists mostly of gas, and roughly $\sim 1$ per cent of dust \citep{CY10_review}. The small dust grains grow into cm-sized pebbles, which later grow into km-sized planetesimals that later form planets.

While the growth up to cm-sized pebbles and the growth of planetesimals into planets are fairly well understood, the formation of the first planetesimals poses a major challenge.  While small grains are tightly coupled to the gas flow and can efficiently grow to mm-cm pebbles, larger $\sim$ metre-sized boulders experience collisional fragmentation and erosion, or interact through bouncing rather than sticking, and are susceptible to strong gas-drag induced radial drift \citep{Wei77}. Such boulders are therefore rapidly lost, not allowing for planetesimal growth beyond these typical sizes.  

Various pathways to overcome the metre-size barrier problem were suggested \citep{CY10_review,Blu18}. These include the gravitational collapse of overdense regions into  large planetesimals, where the overdensity of dust and pebbles is catalyzed by streaming instabilities \citep{Y05,J07}. Other channels involve rare cases of successful collisional growth into large planetesimals under favourable conditions in terms of velocity distribution and/or composition \citep{Win+12a,Boo+18,Blu18}. However, all of these scenarios encounter major challenges and are not robust, as they require highly fine-tuned conditions (see refs.  \citealp{CY10_review,Blu18} for an overview).   
Once planetesimals reach km-size, further growth is achieved by gravitational interactions, and accretion of pebbles is efficient in the presence of massive planetary embryos  \citep{Orm+10,LJ12}. One of the main challenges for planet formation is therefore the initial formation of km-sized planetesimals. 

The recent flyby of the interstellar-object 'Oumuamua \citep{Meech17}
suggests that encounters of interstellar planetesimals with different
solar systems are much more common than previously thought \citep{Do18}.
Such interstellar planetesimals were suggested to be potentially recaptured later-on into other solar-systems through purely dynamical processes \citep{Adams+05,Valtonen09,Lev+10,B12,Perets+12}, but they are inefficient and/or occur at late times after planet-formation processes take-place.

Here we propose that a different efficient \emph{gas-assisted} capture process takes place when a gaseous PPD
 still exists. Gas-dust/planetesimals interactions are known to play an important role in planet-formation and the evolution of bound-planetesimals embedded in PPDs \citep{Ada+76,Wei77,Cuk04,PMC11, Fujita13,GP15}. Small grains and pebbles are decelerated by aerodynamic gas-drag in the disc. Here we show that gas-drag assisted capture of \emph{ interstellar} planetesimals capture is no-less important.

In this paper, we show that planetesimal-capture through this process could play an important role in the initialization and catalysis of efficient planet-formation, thereby  alleviating the metre-size barrier problem and providing a robust mechanism for the initial $\sim$km-size planetesimal seeds needed for efficient planet-formation.  Therefore, gas-assisted capture of interstellar planetesimals can potentially resolve some of the main difficulties in our understanding of planet formation through the provision of planetesimal seeds into young PPD. 
 
Our paper is organized as follows: In sec. \ref{sec2} we estimate the encounter rate of ISM planetesimals from the mass function of ejected planetesimals (sec. \ref{sec21}) and the local stellar environment (sec. \ref{sec22}). In sec. \ref{sec3} we present the analytical planetesimal capture model. We review the PPD structure (sec. \ref{sec31}), derive the capture condition (sec. \ref{sec32}) and the capture rates (sec. \ref{sec33}). We compare our results with Monte-Carlo simulations (sec. \ref{sec34}) and estimate the number of captured objects during the PPD's lifetime (sec. \ref{sec35}). In sec. \ref{sec4} we discuss theimplications and caveats of the model and summarize in sec. \ref{sec5}.

\section{Encounter rate} \label{sec2}

\subsection{Planetesimal mass function}  \label{sec21}
During the planet
formation process, a large amount of planetesimals is ejected from
a given planet-forming system, and these become unbound interstellar-objects \citep{Dones99, Melosh2003}. These ejections occur both during the early planet
formation phase, or on longer timescales throughout the stellar and
dynamical evolution of the system, long after the PPD
dissipates. \cite{Adams+05} estimate that for each young
star at least $\gtrsim M_{\oplus}$ of solids are ejected into the interstellar medium (ISM)
with a typical ejection velocity of $\langle v_{\rm{eject}}\rangle=6.2\pm2.7\ {\rm km/s}$.
They consider a mass function of $dN_{\rm eject}/dm\propto m^{-p}$, whence the total number of ejected planetesimals up to mass $m$  is \citep{Adams+05}
\begin{equation}
N_{\rm eject}(m)=\frac{2-p}{p-1}\frac{M_{T}}{m^{p-1}m_{{\rm up}}^{2-p}} \label{eq:nmgen}
\end{equation}
where $M_{T}$ is the total mass, and $m_{{\rm up}}$ is the upper
cutoff of the largest mass possible. Following \cite{Adams+05} we adopt $m_{{\rm up}}=0.1M_{\oplus}$
and $M_{T}=M_{\oplus}$ (a conservative value).

The power law depends on the details of the formation of the first planetesimals. \cite{Adams+05} use a power-law with $ p = 5/3$, which is also consistent with recent streaming instability (SI) simulations  $p=1.6\pm0.1$ \citep{Simon17}. However, the SI formed planetesimals are too large, ($R  \gtrsim 10\ \rm km$), and it is more reasonable to consider Dohnanyi-like distributions \citep{d69} of collisional cascade, leading to $p=11/6$ \citep{Raymond17}. For $p=11/6$, the number of ejected planetesimals of mass $>m_{1}$ is then  $N_{\rm eject}(m>m_{1})\sim 0.3\left(m_{1}/M_{\oplus}\right)^{-5/6}$. 

Another alternative possibility is that a fraction of interstellar
planetesimals was disrupted during ejection (\citealp{Raymond18} found 0.1-1 per cent.) at preferred radius $r_{{\rm disr}}\approx100{\rm \ m}$.
Given a total mass of $M_{T}\approx10^{-3}M_{\oplus}$, the number
of planetesimals of size $r_{{\rm disr}}$ is $N=M_{T}/(4\pi r_{{\rm disr}}^{3}\rho_{p}/3)\approx10^{12}$,
which is comparable to the number density from the original distribution
$N(m_{{\rm disr}})\approx0.3\cdot10^{12}.$ Thus, the enhancement
is by at most a factor of a few. If there is a distribution of $r_{{\rm disr}}$,
or if $r_{{\rm disr}}$ is increase, the resulting enhancements will
be smaller. 

\subsection{Encounter rates at different environments}  \label{sec22}
The number of planetesimals entering the disc
 region is therefore
$N_{\rm enter}\approx n_{\rm ISM}\sigma_{\rm env}\langle v\rangle \tau_{\rm env}$, where
$n_{\rm ISM}=n_{\star}N_{\rm eject}$ is the number density of interstellar planetesimals,
$n_{\star}$ is the number density of stars, $\sigma_{\rm env}=\pi b_{\rm max}^{2}$
is the cross section, with a maximal impact parameter $b_{\rm max}$, above
which no significant encounter occurs, $\langle v\rangle = \sqrt{8/\pi} \sigma$ is the
mean velocity where $\sigma$ is the velocity dispersion of the environment, and  $\tau_{\rm env}$ is a typical timescale during which encounters with the disc can occur.

One may consider two types of environments;
(1) A cluster/stellar-association environment in which a group of
stars is bound together and their relative velocities are low; and
(2) a field environment where stars and/or interstellar planetesimals are
unrelated to each other and the relative velocities between them are
high. For the cluster environment we consider a stellar density of $n_{\star}^{c}\sim750/\pi N_{\star}^{1/2}\ {\rm pc}^{-3}$,
where $N_{\star}=100-1000$. In this case, the velocity is dominated
by the dispersion velocity of ejected planetesimals $\sigma =6.2\pm2.7\ {\rm km/s}$ \citep{Adams+05}. 
In the field, the velocity is dominated by the (observed) stellar
velocity dispersion $\sigma \sim 30\ \rm km/s$, and the stellar
density is $n_{\star}^{f}\sim0.1\ {\rm pc}^{-3}$. Young systems are likely to form in the central parts of the Galactic disc \citep{tre04}. The velocity dispersion of young stars and stars residing in the central part of the disc is therefore typically lower than assumed here (i.e. $\sim20\ \rm km/s$), and therefore our fiducial choice is likely to be conservative. Moreover, additional environments can be considered, such as globular cluster, and moving stellar group \citep{tre04} that have larger number density or small velocity dispersion, respectively.

For the cluster, $\tau_{\rm env} = r/\sigma \sim 0.3$ Myr is the cluster crossing-time. In the field, the typical time is dominated by the disc lifetime $\tau_{\rm env} = t_{\rm disc}=3$ Myr.
In the case of a young cluster environment the timing of material ejection is important. In particular,  if ejections take place at times much longer than the lifetime of PPDs they will not contribute to the reservoir of interestellar-planetesimals available for capture. However, models suggest the actual timescale for material ejection is comparable to that of the gaseous disc lifetime \citep{Mor18}.  

Plugging in the numbers, the number of planetesimals entering a PPD
during its lifetime is  $N_{\rm enter}^{c}(R>1\ {\rm km})\approx1.1\cdot10^{5}(R/{\rm km})^{-5/2}$
in a cluster environment, and $N_{\rm enter}^{f}(R>1\ {\rm km})\approx10^{4}(R/1\ {\rm km})^{-5/2}$
in the field. This is likely a lower-limit, since the inferred encounter rate of 'Oumuamua-like objects (with effective diameter of $\sim100\ {\rm m}$), given its recent detection, is  $\sim$50 times higher than the above-estimated rate for $100\ \rm{m}$-size
bodies entering the Solar system in today's field environment \citep{Do18}.

\section{capture rates} \label{sec3}

\begin{figure*}
\begin{centering}
\includegraphics[height=6.5cm]{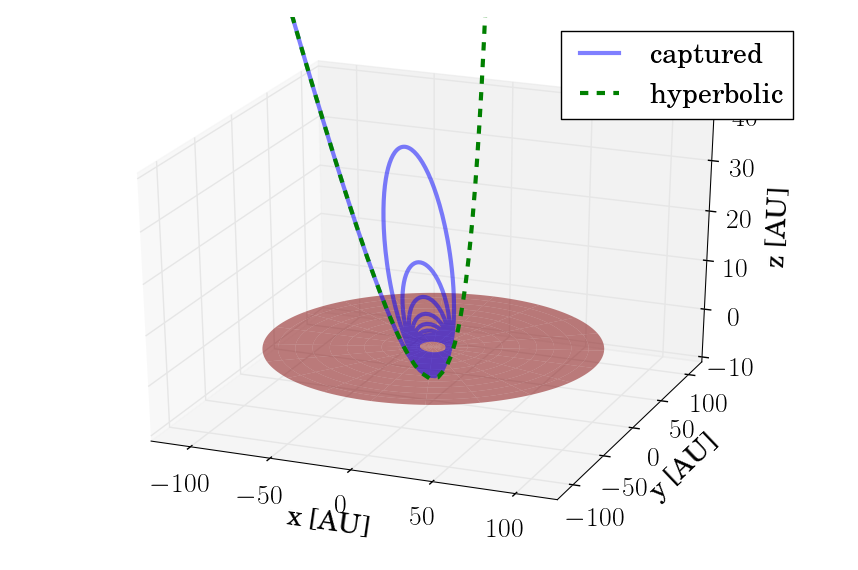}\includegraphics[height=6.5cm]{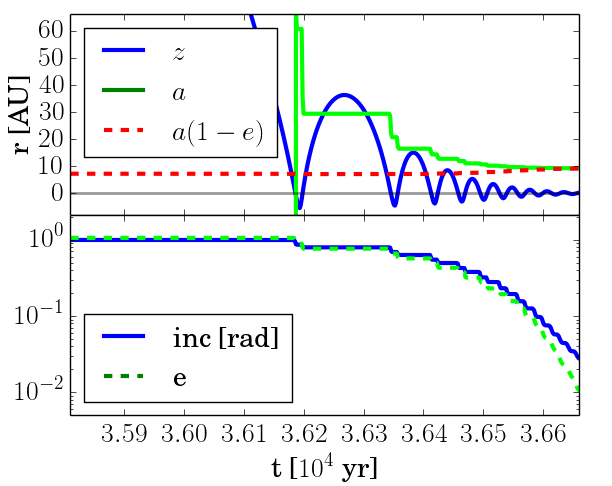}
\caption{\label{fig:1}Flyby and capture orbits of interstellar planetesimals. Left: The 3D trajectories of initially hyperbolic, interstellar planetesimals. The dashed green line represents a hyperbolic (non-capture) encounter, similar case of 'Oumuamua, for planetesimals of size $R_p = 10^4\ \rm m$.  The blue line corresponds to a smaller planetesimal ($R_p = 10\ \rm m$) which efficiently dissipates its energy through gas-drag, decelerates and becomes embedded in the disc (red circle). The initial orbital elements are the same. Right: Time evolution of the orbital elements of the captured orbit. Top panel shows the evolution of the height $z$ above the PPD, the semi-major axis $a$ and the pericentre approach $a(1-e)$. The bottom panel shows the evolution of the inclination and the eccentricity $e$.  }
\end{centering}
\end{figure*}  
 
\subsection{Protoplanetary disc structure}  \label{sec31}
The radial and vertical structure of the PPD can be modeled from the Chiang-Goldreich simple flared disc model \citep{CG97}. The radial  gas surface density is $\Sigma_{g} = \Sigma_{g,0} (a/ {\rm AU})^{-\beta}$, where $\Sigma_{g,0} = 2\cdot 10^3 \rm {g\  cm^{-2}}$. The normalization $\Sigma_0$ and scaling $\beta = 3/2$ corresponds to the Minimal Mass Solar Nebula (MMSN) profile \citep{wei77_mmsn, Hayashi81}. Larger normalizations and various power laws ($\beta \sim 0.5-2.2$) have been invoked in order to explain the formation of Super-Earth planets, though there is a large spread and uncertainty in the observed systems \citep{chiang13_mmsn, raymond14_mmsn}. 

The vertical structure is governed by hydrostatic equilibrium, which leads to a Gaussian profile, where in cylindrical coordinates $\rho_g(r,z) = \rho_g(r,0) \exp(-z^2 / 2 h^2)$, where $h = c_s / \Omega$ is the disc scale height, $c_s$ and $\Omega$ is the sound speed and the Keplerian frequency, respectively \citep{PMC11, GP15}. The surface density is then $\Sigma_g (r) =\int \rho_g (r,z) dz = \sqrt{2 \pi} h \rho(r,0)$. 

\subsection{Capture condition} \label{sec32}
 
Consider an interstellar object coming from infinity with velocity $\boldsymbol{v}_{\infty}$, going through a gaseous PPD around a star of mass $M_{\star}$. For a spherical body with density $\rho_{p}=1\ {\rm g\ cm^{-3}}$, radius $R_{p}$ and relative velocity $\boldsymbol{v}_{\rm rel}$ which crosses a region of the disc with with density $\rho_{g}$, the
aerodynamic gas drag force is

\begin{equation}
\boldsymbol{F}_{\rm D}=-\frac{1}{2}C_{D}\pi R_p^{2}\rho_{g}v_{\rm rel}^{2}\hat{\boldsymbol{v}}_{\rm rel},\label{eq:drag}
\end{equation}
where $C_{\rm D}(\pazocal{R}e)$ is the drag coefficient, which depends on the Reynolds number $\pazocal{R}e=\sim R_{p}v_{\rm rel}/\nu_{m}$, where $\nu_{m}=(1/2)\bar{v}_{\rm th}\lambda$ is the molecular viscosity of the gas, $\bar{v}_{\rm th}$ is the thermal velocity and $\lambda	$ is the mean free path of gas-gas collisions.
 
Large planetesimals are in the ram pressure regime with constant coefficient $C_{\rm D}=0.44$. Small dust grains are in the Epstein regime, with $C_{\rm D} \propto R_p^{-1}$. The transition to Stokes regime occurs at $R_p = 9\lambda/4$. In the Stokes regime, $C_{\rm D} \propto R_p^{-3/5}$.  We follow \cite{PMC11} for prescription for the Reynolds number and drag laws. 

For a planetesimal that crosses the disc face on at radial location $\bar{a}$, the amount of energy
loss during the interaction with the disc is the total work applied
on the planetesimal 
\begin{align}
\Delta E & = \intop\boldsymbol{F}_{D} \cdot \boldsymbol{v}_{\rm rel} {\rm d}t=-\frac{\pi C_{D} R_{p}^{2}\rho_0 }{2} \intop_{-\infty}^{\infty}\exp\left(-\frac{v_{\rm rel}^{2}t^{2}}{2h^{2}}\right)v_{\rm rel}^{3}{\rm d}t \nonumber \\
 & \approx -\frac{\pi C_{D}}{2} R_{p}^{2}\Sigma_{g}v_{\rm rel}^{2}, \label{de}
\end{align}
where $\rho_0 = \rho_g(\bar{a},z=0)$ is the density at the midplane, and we assume that the relative velocity $\boldsymbol{v}_{\rm rel}=\boldsymbol{v}_{\infty}+\boldsymbol{v}_{\rm esc}(\bar{a})$ is constant throughout the passage, where $v_{\rm esc} = \sqrt{2 G M_{\star} / \bar{a}}$ is the escape velocity.

There are two regimes: Either $v_{\infty} \gg v_{\rm esc}$, the geometrical regime, or $v_{\infty} \ll v_{\rm esc}$, gravitational focusing regime, letting  $v_{\rm rel}^{2}\approx v_{\infty}^{2}+v_{\rm esc}^{2}$ takes both options into account. 

The body is captured if it has dissipated more energy than its initial energy  $E_{\rm in}=m_{p}v_{\infty}^{2}/2$, or $|\Delta E| > E_{\rm in}$. In terms of the body's size, the capture condition is

\begin{equation}
R_{p}\lesssim\frac{3}{4}\frac{C_{\rm D}\Sigma_{g}}{\rho_{p}}\left(1+\Theta_{s}\right),\label{eq:capture_cond}
\end{equation}
where $\Theta_{s}\equiv v_{\rm esc}^{2}/v_{\infty}^{2}$ is the gravitational
focusing Safronov number. For $\Theta_{s}\gg1$
gravitational focusing is important, while for $\Theta_{s}\ll1$ the
scattering is mostly in the geometric collision regime. Intuitively,  in the geometric regime, capture requires a velocity change of order of the incident velocity $\delta v \sim v_{\rm \infty}$, thus corresponds to a requirement that the disc surface density exceeds a fixed fraction of the planetesimal mass per unit area.

  Fig. \ref{fig:1} shows an example of a typical trajectory and evolution of interstellar planetesimals as they encounter the disc,
dissipate their energy and become embedded in the disc. In the following
we use a detailed analysis to provide a quantitative study of the
capture rate of such planetesimals.

\subsection{Capture rates}  \label{sec33}

In order to evaluate the fractions and total number of planetesimals
captured through this process, we need to consider the properties of the orbits and the PPD, as well as properties of the environment. 

For the encounter properties, we consider the distributions of the velocity, impact parameters and relative impact angles to the disc, as well as the size-distribution
of the incoming planetesimals. We assume that Interstellar-objects have a Maxwellian velocity distribution (similar to their progenitor stellar hosts)
\begin{equation}
f_{V}(v_{\infty},\sigma)  =  \sqrt{\frac{2}{\pi}}\frac{v^{2}}{\sigma^{3}}\exp\left(-\frac{v_{\infty}^{2}}{2\sigma^{2}}\right);\ v\in[0,\infty], \label{eq:fv}
\end{equation}

where $\sigma$ is the velocity dispersion. Moreover, faster planetesimals collide more frequently so the distribution of rate of collisions per unit time is further weighted by an additional factor of $v_{\infty}$. The distribution of impact
parameters follows a simple geometric cross-section, i.e. a
uniform distribution of the impact parameter $B^{2}\sim U[0,b_{\rm max}^{2}]$
(the trajectory can later change due to gravitational focusing, which
we account for when relevant), where $b_{\rm max}$ is the maximal impact
parameter for an effective close encounter. Both of these depend on
the stellar environment.

\subsubsection{Geometric regime}

In this case $\Theta_{s}\ll1$ (negligible gravitational focusing), i.e. the trajectory of an incoming interstellar planetesimals follows a straight line before encountering the disc, and is negligibly affected by the gravitational pull from the host star. The capture criteria is then
\begin{equation}
\frac{3C_{\rm D}\Sigma_{g}}{4\rho_{p}R_{p}}<1
\end{equation}
taking a density profile $\Sigma_{g}=\Sigma_{g,0}(a/{\rm AU})^{-\beta}$
we get 
\begin{equation}
\left(\frac{a}{{\rm AU}}\right)^\beta<\frac{3C_{\rm D}\Sigma_{g,0}}{4\rho_{p}R_{p}}
\end{equation}

For geometric scattering, the closest approach is $q\approx b$, so
the criteria is 
\begin{equation}
b_c(R_p)<\left(\frac{3C_{\rm D}\Sigma_{g,0}}{4\rho_{p}R_{p}}\right)^{1/\beta}{\rm AU}
\end{equation}

or with the dimensionless parameter $x=b_c/b_{\rm max},$ the capture probability
is 
\begin{align}
f_c(R_p) & = \left(\frac{b_c}{b_{\rm max}}\right)^{2}=\left(\frac{3C_{\rm D}\Sigma_{g,0}}{4\rho_{p}R_{p}}\right)^{2/\beta}\left(\frac{{\rm AU}}{b_{\rm max}}  \right)^{2} \nonumber \\
 & = 0.16\left(\frac{C_{\rm D}}{25}\right)^{4/3}\left(\frac{b_{\rm max}}{130\ {\rm AU}}\right)^{-2}\left(\frac{R_{p}}{2\ {\rm m}}\right)^{-4/3},\label{eq:fc_geometric}
\end{align}

\begin{figure*}
\includegraphics[height=6.5cm]{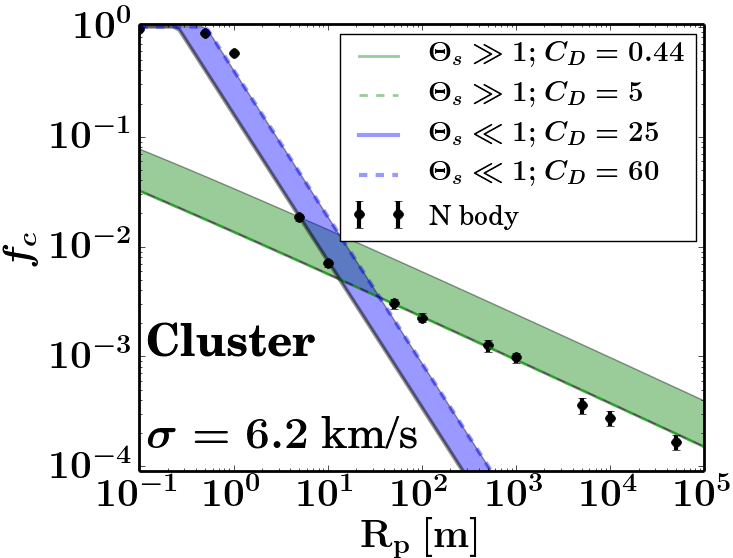}\includegraphics[height=6.5cm]{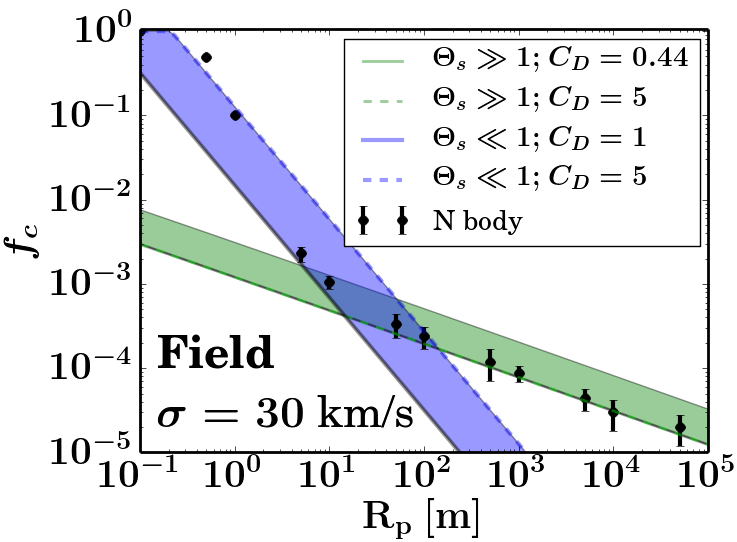}\caption{\label{fig:2}Fractions of captured planetesimals as a
function of their size. Blue area is the estimated probability in
the geometric scattering regime, green area is the estimated probability
in the gravitational focusing regime. Black dots are result of a numerical simulation.
The error bars are estimated from shot noise of the captured orbits. Choices of drag coefficients are discussed in the Methods.
Generally, planetesimals above $R_{p}\gtrsim100\ \rm m$ follow the
gravitational focusing prediction $f_{c}\propto R_{p}^{-2/5}.$ The
transition to geometric scattering regime is near $\sim10\  \rm m$
for both environments $\sim100\ {\rm m}$.}
\end{figure*}

where we used $\beta = 3/2$ and the disc normalization of sec. 2.1. The geometric regime is independent of the initial velocity and the PPD profile. The geometric regime is valid mostly for small grains and pebbles, white neglecting gravitational focusing under-predicts the capture probability of large $\gtrsim 100 \rm \ m$ planetesimals.

\subsubsection{Gravitational focusing regime}

 In the gravitational focusing regime, ($\Theta_{s}\gg1$;
which corresponds to small impact parameters and/or low velocities), the capture condition is 
\begin{equation}
\left(\frac{a}{{\rm AU}}\right)^{\beta}v_{\infty}^{2}<\frac{3C_{\rm D}\Sigma_{g,0}}{2\rho_{p}R_{p}}\frac{GM}{a},\label{eq:gfoc}
\end{equation}
which depends both on the velocity and the impact parameter. 
In order to proceed, we use the parabolic approximation to find the
closest approach $a=q$,
\begin{equation}
q=\frac{GM}{v_{\infty}^{2}}\left[\sqrt{1+\frac{b^{2}v_{\infty}^{4}}{G^{2}M^{2}}}-1\right]\approx\frac{b^{2}v_{\infty}^{2}}{2GM}
\end{equation}
so the capture condition is 
\begin{equation}
b^{2(1+\beta)}\le2^{1+\beta}\frac{3C_{\rm D}\Sigma_{g,0}}{2\rho_{p}R_{p}}\left(\frac{GM}{v_{\infty}^{2}{\rm AU}}\right)^{2+\beta}{\rm AU}{}^{2(1+\beta)}\equiv b_c^{2(1+\beta)}\label{eq:15}
\end{equation}
The velocity dependent capture probability is 
\begin{equation}
P_{c}(R_{p}|v_{\infty})  =  \intop_{0}^{b_{c}(v_{\infty})}f_{B}(b){\rm d}b=\left(\frac{b_{c}(v_{\infty})}{b_{{\rm max}}}\right)^{2} \label{eq:pcvinf}
\end{equation}
where $b_c(v_{\rm \infty})$ is given by equality in Eq. (\ref{eq:15}). Now we want to look at the different rates of arrival: faster planetesimals
have higher encounter rates than slower one. The integrals that involve
the probability have an additional $v$ factor, e.g. $P\propto\intop vf_{V}(v){\rm d}v$.
Namely, the weighted capture probability for a given time is 
\begin{align}
f_c(R_{p}) & = \frac{N_{{\rm captured}}(R_{p})}{N_{{\rm enter}}(R_{p})}\nonumber \\
& = \frac{\intop N_{{\rm enter}}(v,R_{p},b_{{\rm max}})f_{V}(v_{\infty})P_{c}(R_{p}|v_{\infty}){\rm d}v_{\infty}}{\intop N_{{\rm enter}}(v,R_{p},b_{{\rm max}})f_{V}(v_{\infty}){\rm d}v_{\infty}} \label{p1}
\end{align}
Since $N_{{\rm enter}}(R_{p})\propto v_{\infty}$ we have 
\begin{align}
f_c(R_{p}) & =\frac{1}{\langle v_{\infty}\rangle_{f_{V}}}\intop v_{\infty}f_{V}(v_{\infty})P_{c}(R_{p}|v_{\infty}){\rm d}v_{\infty} \nonumber \\
& = \iint_{D_c} \frac{2b}{b_{\emph{\rm max}}^{2}}\frac{v^{3}}{2\sigma^{4}}e^{-v^{2}/2\sigma^{2}} {\rm d}b{\rm d}v \label{p2}
\end{align}
 where $\langle v_{\infty}\rangle_{f_{V}}=\sqrt{8/\pi}\sigma$ is
the mean thermal speed. The integration is on the fractional domain $D_c$ of parameters that result in capture. 

The latter is equivalent to drawing the velocity from a $\chi(4)$ distribution:
\begin{equation}
\tilde{f}_{V}(v)=\frac{1}{2}\frac{v^{3}}{\sigma^{4}}e^{-v^{2}/2\sigma^{2}}
\end{equation}

The capture probability is encapsulated in the new random variable $x\equiv(b/b_{\rm max})^{\alpha}(v_{\infty}/\sigma)^{2}$, where $\alpha = 2 (1+\beta)/(2+\beta)$. We show in Appendix A that the distribution function for $x$ can be expressed in terms of incomplete Gamma functions,

\begin{align}
f_{X}(x) = 2^{-(2+\beta)/(1+\beta)}\frac{2+\beta}{1+\beta}\Gamma\left(\frac{\beta}{1+\beta},\frac{x}{2}\right)x^{1/(1+\beta)}\label{eq:fx}
\end{align}
For the special case of $\beta=3/2$, $ f_{X}(x) \propto x^{2/5}\Gamma\left(\frac{3}{5},\frac{x}{2}\right)$,
and the weighted capture probability is 
\begin{align}
f_c(R_{p}) & =\intop_{0}^{x_c(R_p)}f_{X}(x){\rm d}x \nonumber \\
x_c & =2^{\beta/(2+\beta)}\left(\frac{3C_{\rm D}\Sigma_{0}}{\rho_{p}R_{p}}\right)^{1/(2+\beta)}\left(\frac{GM}{\sigma^{2}{\rm AU}}\right)\left(\frac{b_{{\rm max}}}{{\rm AU}}\right)^{-\alpha(\beta)}\label{eq:pgf}.
\end{align}

Since the gravitational focusing regime is relevant for large planetesimals, it required close approach, which is possible only with either small velocity of impact parameter, or $x_c \ll 1$. In this case, the (weighted) probability (Eq. \ref{eq:pgf}) can be expanded into leading terms,
\begin{equation}
f_c(x_c)  = \left(\frac{x}{2}\right)^{(2+\beta)/(1+\beta)} \Gamma\left(\frac{\beta}{1+\beta}\right) + \pazocal{O}(x_c^2).\label{eq:cdf2}
\end{equation}
For the special case $\beta = 3/2$ we have
\begin{align}
f_c(R_p) & =\frac{\Gamma\left(\frac{3}{5}\right)}{2^{7/5}}x_c^{7/5} + \pazocal{O}(x^2_c) \nonumber \\
 & \approx0.86\left(\frac{3C_{\rm D}\Sigma_{0}}{\rho_{p}R_{p}}\right)^{2/5}\left(\frac{GM}{\sigma^{2}{\rm AU}}\right)^{7/5}\left(\frac{b_{{\rm max}}}{{\rm AU}}\right)^{-2}. \label{eq:prob}
\end{align}
Note that in either regime of the capture fraction (Eq. \ref{eq:fc_geometric}
and \ref{eq:prob}) it is proportional to $\propto b_{\rm max}^{-2},$
which cancels out with the $\propto b_{\rm max}^{2}$ from the encounter
rate. Thus, the total number of captured planetesimals $N_{\rm enter}\cdot f_{c}$
is independent of $b_{\rm max},$ hence the choice of $b_{\rm max}$ is rather
arbitrary, as expected. Multiplying by $n_{{\rm ISM}}(\sqrt{8/\pi}\sigma)\pi b_{{\rm max}}^{2}$
yields the total number captures: 
\begin{equation}
N(R_{p}) \approx 4.29 \left(\frac{3C_{\rm D}\Sigma_{0}}{\rho_{p}R_{p}}\right)^{2/5}\left(\frac{GM}{\sigma^{2}{\rm AU}}\right)^{7/5}{\rm n_{{\rm ISM}}\sigma\tau_{{\rm env}}AU^{2}},
\end{equation}
and it is independent of $b_{\rm max}$, as expected.
\subsection{Numerical modeling and comparison}  \label{sec34}

In order to better verify the analytical estimates,
we run N-body simulations that include gravity and a prescription
for gas drag (Eq. \ref{eq:drag}), based on 4th order Hermite integrator \citep{Hut95}. We truncate the disc density at $r_{\rm disc} = 250\  \rm AU$. The aspect ratio is $h/r = 0.022 (r/ \rm AU) ^ {2/7}$. The velocity of the gas is slightly sub-Keplerian due to pressure gradients, namely $v_{\rm gas} = \eta v_{\rm Kep}$, with $\eta = (1 - (3/2 + \beta + 3/14) (h/r)^2)^{1/2}$ (cf. \citealp{GP15, GP16} for more details). The relative velocity is $\boldsymbol{v}_{\rm rel} = \boldsymbol{v}_{\rm p} - \boldsymbol{v}_{\rm gas}$, where $\boldsymbol{v}_{\rm p}$ is given in cartesian coordinates after rotation of the hyperbolic orbit to the disc's reference plane. We initialize the planetesimal to start from $r_0 = 20000 \ \rm AU$ with orbital parameters and disc inclination drawn from distributions described in the main text. We stop the simulation if the distance from the sun exceeds $50000\  \rm AU$ and negative energy and conclude the orbit is unbound. For bound orbits we stop if either the distance is $r < 0.02\  \rm AU$ or the orbital eccentricity is $e<0.1$. 

 For each planetesimal-size we run $10^{4}-3\cdot10^{5}$
numerical integrations with $b$ and $v_{\infty}$ distributed from uniform in $b^2$ and  $\chi(4)$ distributionm distribution, respectively. The relative angles between the planetesimal trajectories and
the PPD were drawn from an isotropic distribution
(uniform in the argument of pericentre and the longitude of ascending node angles, and uniform in the cosine of the inclination
angle). 

Fig. \ref{fig:2} shows the comparison between the analytic
estimates and the simulations. Small pebbles, up to $\sim1\ {\rm m}$, are the most susceptible to gas drag, and are efficiently captured.
These can be captured even at the lower density regimes of the disc
at large separations. Larger planetesimals require progressively close pericentre approach of their trajectory, near the high-density inner regions of the disc. Therefore, large planetesimals are in the gravitational focusing regime, where the gas drag is ram-pressure dominated and the drag coefficient has a constant value $C_{\rm D}=0.44$. For geometric scattering we expect to be somewhere near the Epstein-Stokes transition, i.e. near $C_{\rm D}=24$.

Fig. \ref{fig:2} shows good converge of the analytical models, both in the power law scaling and in the gas drag regime. Planetesimals in the field have much higher velocities and reynolds numbers, hence they are rarely in the Epstein regime and have smaller drag coefficients.

\subsection{Total number of captured planetesimals and radial distribution}  \label{sec35}

\begin{figure*}
\begin{centering}
\includegraphics[width=8.5cm]{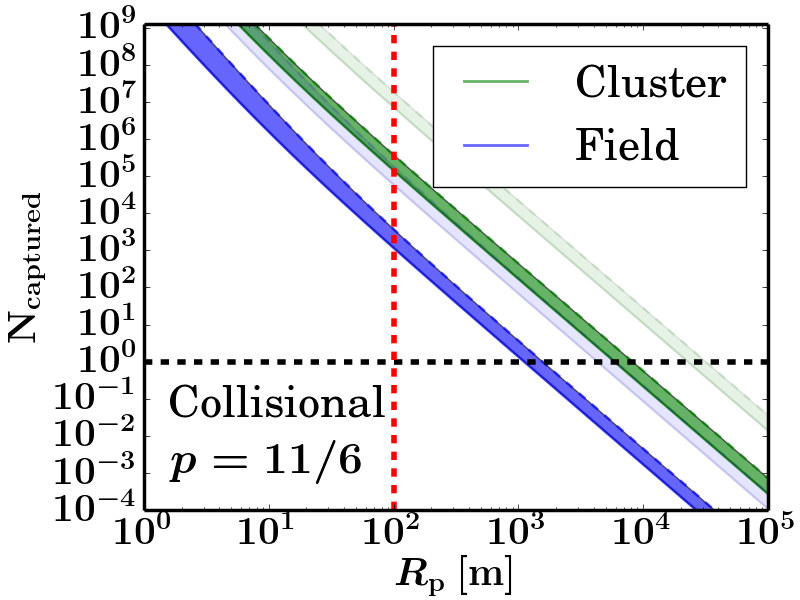} \includegraphics[width=8.5cm]{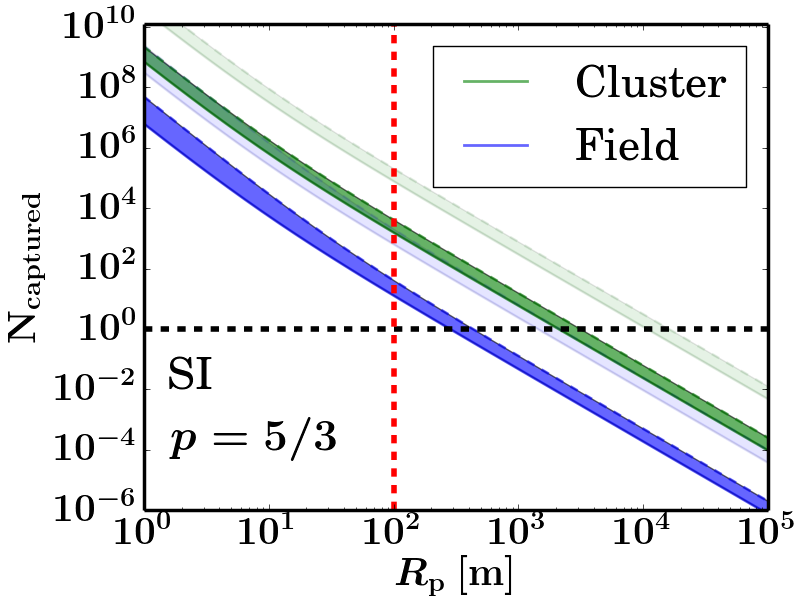}
\caption{\label{fig:Ncaptured} The number of captured planetesimals during a typical
lifetime $\tau_{\rm env}$, given the theoretically estimated
rates. The left and right panels represent collisional (Dohnanyi) and SI induced mass functions. Green area corresponds to the rates for the cluster, while blue area corresponds to the rates for the field. The boundaries are determined by different drag coefficients, similar to Fig. \ref{fig:2}.  Red vertical line stands for the effective radius of 'Oumuamua, $\sim 100\ \rm m$. The observed rate, based on 'Oumuamua passage, are enhanced by $\sim50$ times. The enhanced rates represented by the transparent green and blue areas for the cluster and the field, respectively.}
\end{centering}
\end{figure*}

Using the size-dependent capture-probability, we obtain the total
number of captured planetesimals of a given size. 

The left panel of Fig. \ref{fig:Ncaptured} shows the expected size-dependent number
of captured planetesimals for a collisional planetesimal mass function of $p=11/6$. Many small pebbles and planetesimals up to $\lesssim 100 \ \rm m$ are captured, which could lead to efficient seeding and subsequent planet formation. At least one planetesimal as large as $\sim6\ {\rm km}$ ($\sim1\ {\rm km})$ is captured in a cluster (field) environment. The inferred rate, based on 'Oumuamua passage, is enhanced by $\sim50$ times. The latter would then result in the capture of even $\sim23\ {\rm km}$ ($\sim4\ {\rm km})$ for a cluster (field) environment. The right panel of Fig. \ref{fig:Ncaptured} shows the the number of captured planetesimals from SI mass function, $p=5/3$. It this case, there are less planetesimals to begin with, therefore the overall numbers are lower, althoguh still significant.

\begin{figure*}
\includegraphics[height=6.3cm]{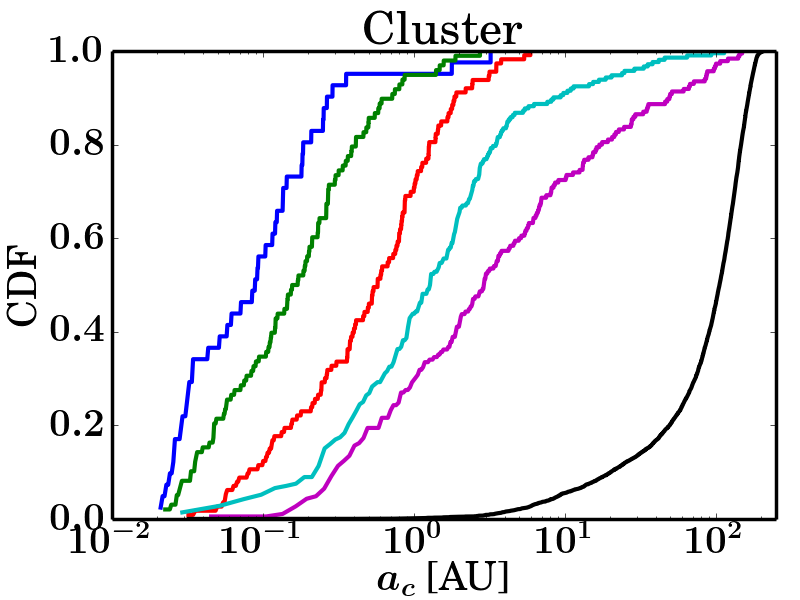}\includegraphics[height=6.3cm]{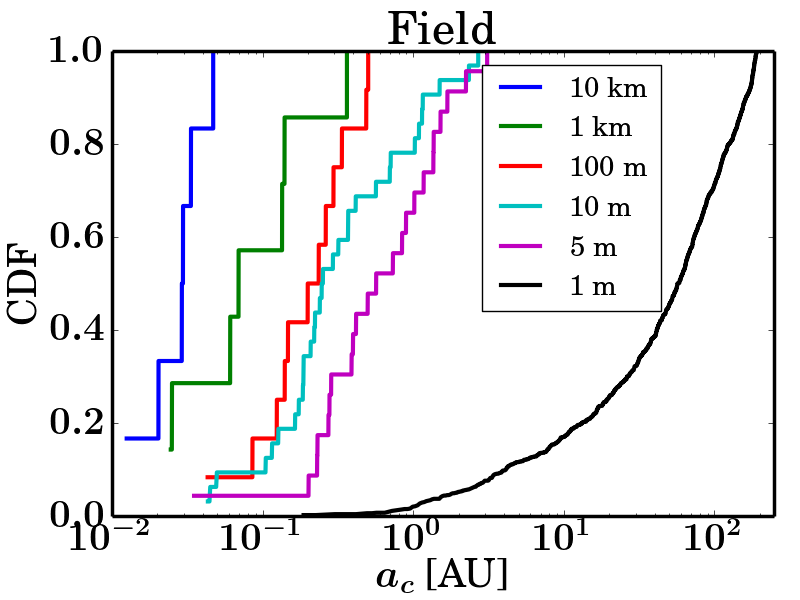}
\caption{\label{fig:radial-distribution}The size dependent cumulative radial
distribution of captured planetesimals. Several planetesimal sizes
are shown; $R_{p}=10^4,\ 10^3\ ,100,\ 10\ ,5\ ,1\  {\rm m}$ (blue, green, red, cyan, purple and black respectively), for the cluster
(left) and the field (right). Once captured, the circularization timescale is much shorter than radial drift timescales (see Fig. \ref{fig:1}b and \protect\cite{Ada+76} ), thus the final semi-major axis is close to the first pericentre approach.}
\end{figure*}

Fig. \ref{fig:radial-distribution} shows the empirical cumulative radial distribution of captured planetesimals for different size ranges.
As mentioned above, disc dissipation can be efficient for small planetesimals
even at the disc outskirts where the gas densities are low. The capture of larger planetesimals, however,  requires higher gas densities. Therefore the larger the planetesimals, the more centrally concentrated is their radial distribution.  

\section{Discussion}  \label{sec4}

\paragraph*{The meter size barrier and the first planetesimals:}
The gas-assisted capture mechanism can seed a few up to thousands of $\sim \ \rm km$-sized planetesimals in the disc. Such relatively small number of seeds can then rapidly grow to $100\ \rm km$-size on short time-scales before collisional erosion starts \citep{Xie+18}. We therefore expect the effective initial distribution of cores and asteroids to be at large radii, somewhat similar to that expected from the SI models and consistent with suggestions and observations that asteroids were born big and that the asteroids were formed from a small number of asteroid families \citep{Mor+09,Der+18}.  A fraction of these grown planetesimals later ejected from the systems, and further replenish the population
of interstellar planetesimals. These, in turn,  can be recaptured by other systems and further catalyze planet-formation, and so on, i.e. leading to a chain-reaction -- like exponential planet-seeding process.

One may still question how did large planetesimals and later planets formed in the first system that initialized the seeding. Formation of km-sized planetesimals is a long-standing problem in planet-formation theories \citep{CY10_review,Blu18}. The disc capture scenario can not account for this initial formation, however, it can alleviate the problem, by allowing for the first formation of such planetesimals
to be a rare event, and even under fine-tuned condition. 

Most stars are thought to form in stellar associations
and clusters. If only a few large $\sim 1$ km planetesimals are
sufficient for seeding a given planetary system \citep{Ormel+12, Levison15}, or a larger number of smaller planetesimals \citep{Boo+18},
then $N_{\rm captured}(>1\ {\rm km}) \sim 170$
 large planetesimals are captured by a given system (and numerous smaller ones). Thus, only a small fraction of planetary systems is needed in order to seed all of the
other protoplanetary systems in the cluster ($\sim 0.5\%$ with few big planetesimals, or a negligible fraction for smaller planetesimals). In addition, using the  $\sim50$ times higher rates directly inferred from 'Oumuamua detection, a much smaller number is required, and even a field environment could be seeded by older planetesimals formed in the cluster. We finally note that in the optimistic cases, even large planetary embryos of $10-30$ km can be captured directly.

\begin{table}
\begin{centering}
\begin{tabular}{|c|c|c|c|c|}
\hline 
Cluster & $R_{p}=1{\rm m}$ & $R_{p}=10{\rm m}$ & $R_{p}=10^{2}{\rm m}$ & $R_{p}=10^{3}{\rm m}$\tabularnewline
\hline 
\hline 
Ice & $<1\%
$ & $37\%
$ & $31\%$ & $32\%
$\tabularnewline
\hline 
Carbonate & $<0.1\%
$ & $25\%
$ & $19\%$ & $16\%$\tabularnewline
\hline 
Rock & $<0.1\%
$ & $16\%$ & $10\%$ & $6\%^*$\tabularnewline
\hline 
Iron & $<0.1 \%
$ & $9\%$ & $<6\%$ & $<1\%^*$\tabularnewline
\hline
\noalign{\vskip 4mm}    
\hline  
Field & $R_{p}=1{\rm m}$ & $R_{p}=10{\rm m}$ & $R_{p}=10^{2}{\rm m}$ & $R_{p}=10^{3}{\rm m}$
\tabularnewline 
\hline 
Ice & $7\%
$ & $78\%
$ & $66\%$ & $57\%^*$\tabularnewline
\hline 
Carbonate & $3.7\%
$ & $69\%$ & $33\%$ & $28\%^*$\tabularnewline
\hline 
Rock & $1\%
$ & $53\%$ & $16\%$ & $28\%^*$\tabularnewline
\hline 
Iron & $0.6\%
$ & $34\%$ & $8\%$ & $14\%^*$\tabularnewline
\hline 
\end{tabular}
\par\end{centering}
\caption{\label{tab1} Fraction of captured planetesimals lost due to ablation in field environments. Starred (*) entries are uncertain due to very low or none capture below the ablation limit.}
\end{table}

\paragraph*{Planetesimal ablation:} 
Planetenesimal ablation could potentially affect, or even destroy a planetesimal as it crosses a PPD for the first-time at high velocity. However, using simplified ablation models \citep[Appendix B]{Ar16} we find the expected ablation rates of the captured planetesimals in table \ref{tab1}.  Most $\sim 37\%$ of icy-planetesimals and $\sim 16\%$ of rocky-planetesimals of $10 - 10^3\ \rm m$ sizes are ablated during crossings in stellar clusters, and  $\sim 78\%$ icy and $\sim 53\%$ rocky planetesimals of $10 - 10^3\ \rm m$ are ablated in field environments. Most of the ablated objects are in the range of $10-10^3 \ \rm m$; smaller pebbles are captured far enough, while larger bodies are harder to ablate. These do not significantly affect the overall results.

\paragraph*{Chemical composition:} The composition of meteorites in the Solar System is typically thought
to relate to the primordial composition of the PPD.
However, there is evidence for the existence of material captured
from external sources. In particular, there is evidence for short-lived radioactive nuclei which were likely formed in a relatively nearby supernovae \citep{Ouellette2010}. In addition, analysis of heavy $^{60}\rm Fe - ^{60}\rm Ni$ isotopes in asteroids suggests the early injection of $^{60}\rm Fe$ into the primordial PPD of our Sun \citep{Bizzarro07}. The disc-capture mechanism allows for embedding such
external material in the disc, and may help explain its origin. 

These issues are not the main focus of our study, but it is interesting
to consider the possibility of composition peculiarities in some meteorites
originating from capture of material from other systems through this process. In particular, discovery of rocky/solid material older than the Solar system can provide a signature of material exchange.   

\paragraph*{Lithopanspermia:} The exchange of rocky material between planetary systems may also be considered the leading mechanism for lithopanspermia \citep{Napier,Wes+10,Loeb18}. Even if a small fraction of biologically active material is ejected by interstellar planetesimals, the large efficiency of gas assisted of $1\ \rm m$ sized rocks  could be far more efficient (as much as $\ge10^{5}$ larger) than previously suggested lithopanspermia mechanisms \citep{Adams+05}. 

\section{Summary}  \label{sec5}

 The current paradigm for planet formation involves a bottom up evolution of dust grains growing into planetesimals, then planetary embryos and finally into planets in gaseous protoplanetary discs (PPDs) around young stars.  One of the main open questions over the last forty years deals with the early stages of forming km-size planetesimals from initially small dust grains \citep{wei77_mmsn,CY10_review}. Metre-sized pebbles experience a both significant gas-drag and collisional erosion, and thus rapidly inspiral onto the star or fragment, respectively. Together, these issues give rise to the 'metre-size barrier' problem in which the lifetime of small pebbles in the disc is too short compared with their growth rate. Thus, planet formation requires a 'jump' over small planetesimals as to already begin with larger-size planetesimals not susceptible to these growth constraints. 
	
In this paper we have shown that the majority of planetary systems could have been 'seeded' by large $\sim$ km size interstellar planetesimals. These could have formed elsewhere and then captured via gas-drag experienced by the planetesimal as it passes through the PPD. This gives rise to early insertion of large planetesimals, and it thereby potentially alleviates the meter-size barrier.  The gas assisted capture model does not account for the first generation of planetesimals, but rather makes their formation into a much easier exponentially small challenge. Planetesimal formation is no longer required to form in every planetary system in isolation, but rather in a small subset of systems, perhaps under fine tunes conditions. Even one successful planetesimal formation per system could populate the entire young stellar cluster with planetesimals, and perhaps even the young systems in the field.

We present a novel model of gas assisted capture of interstellar planetesimals. We construct a robust, analytical model of the capture probability and overall capture rates as a function of the planetesimal size and orbit distribution, the protoplanetary disc structure and the local environment. We verified the analytical derivation with direct Monte-Carlo integrations and found good correspondence.

The gas assisted capture model is compatible with late stage planetesimal growth models (snowball phase, pebble accretion, \citealp{Xie+18,LJ12}) and provides the missing link to the initial population of large dust grains and small planetesimals. In addition, the gas assisted capture model supports the observation that asteroids formed big from a small number of asteroid families. The capture model can be tested by future measurements of composition peculiarities in some meteorites, which already has some evidence for early injection of heavy radioactive nuclei into the primordial protoplanetary disc of our Sun \citep{Bizzarro07, Ouellette2010}.

Besides the importance for planet formation, the gas-assisted capture scenario allows for far more efficient exchange of biologically-active material between different planetary systems. In fact, it could be as much as a  million times more efficient than previously estimated \citep{Adams+05}, making the possibility for such panspermia events into the Solar System and/or between other planetary systems far more likely.

\bibliography{drag}

\begin{thebibliography}{}
\makeatletter
\relax
\def\mn@urlcharsother{\let\do\@makeother \do\$\do\&\do\#\do\^\do\_\do\%\do\~}
\def\mn@doi{\begingroup\mn@urlcharsother \@ifnextchar [ {\mn@doi@}
  {\mn@doi@[]}}
\def\mn@doi@[#1]#2{\def\@tempa{#1}\ifx\@tempa\@empty \href
  {http://dx.doi.org/#2} {doi:#2}\else \href {http://dx.doi.org/#2} {#1}\fi
  \endgroup}
\def\mn@eprint#1#2{\mn@eprint@#1:#2::\@nil}
\def\mn@eprint@arXiv#1{\href {http://arxiv.org/abs/#1} {{\tt arXiv:#1}}}
\def\mn@eprint@dblp#1{\href {http://dblp.uni-trier.de/rec/bibtex/#1.xml}
  {dblp:#1}}
\def\mn@eprint@#1:#2:#3:#4\@nil{\def\@tempa {#1}\def\@tempb {#2}\def\@tempc
  {#3}\ifx \@tempc \@empty \let \@tempc \@tempb \let \@tempb \@tempa \fi \ifx
  \@tempb \@empty \def\@tempb {arXiv}\fi \@ifundefined
  {mn@eprint@\@tempb}{\@tempb:\@tempc}{\expandafter \expandafter \csname
  mn@eprint@\@tempb\endcsname \expandafter{\@tempc}}}

\bibitem[\protect\citeauthoryear{{Adachi}, {Hayashi}  \& {Nakazawa}}{{Adachi}
  et~al.}{1976}]{Ada+76}
{Adachi} I.,  {Hayashi} C.,   {Nakazawa} K.,  1976, \mn@doi [Progress of
  Theoretical Physics] {10.1143/PTP.56.1756}, \href
  {http://adsabs.harvard.edu/abs/1976PThPh..56.1756A} {56, 1756}

\bibitem[\protect\citeauthoryear{{Adams} \& {Spergel}}{{Adams} \&
  {Spergel}}{2005}]{Adams+05}
{Adams} F.~C.,  {Spergel} D.~N.,  2005, \mn@doi [Astrobiology]
  {10.1089/ast.2005.5.497}, \href
  {http://adsabs.harvard.edu/abs/2005AsBio...5..497A} {5, 497}

\bibitem[\protect\citeauthoryear{{Belbruno}, {Moro-Mart{\'{\i}}n}, {Malhotra}
  \& {Savransky}}{{Belbruno} et~al.}{2012}]{B12}
{Belbruno} E.,  {Moro-Mart{\'{\i}}n} A.,  {Malhotra} R.,   {Savransky} D.,
  2012, \mn@doi [Astrobiology] {10.1089/ast.2012.0825}, \href
  {http://adsabs.harvard.edu/abs/2012AsBio..12..754B} {12, 754}

\bibitem[\protect\citeauthoryear{{Bizzarro}, {Ulfbeck}, {Trinquier}, {Thrane},
  {Connelly}  \& {Meyer}}{{Bizzarro} et~al.}{2007}]{Bizzarro07}
{Bizzarro} M.,  {Ulfbeck} D.,  {Trinquier} A.,  {Thrane} K.,  {Connelly} J.~N.,
    {Meyer} B.~S.,  2007, \mn@doi [Science] {10.1126/science.1141040}, \href
  {http://adsabs.harvard.edu/abs/2007Sci...316.1178B} {316, 1178}

\bibitem[\protect\citeauthoryear{{Blum}}{{Blum}}{2018}]{Blu18}
{Blum} J.,  2018, preprint, \href
  {http://adsabs.harvard.edu/abs/2018arXiv180200221B} {} (\mn@eprint {arXiv}
  {1802.00221})

\bibitem[\protect\citeauthoryear{{Booth}, {Meru}, {Lee}  \& {Clarke}}{{Booth}
  et~al.}{2018}]{Boo+18}
{Booth} R.~A.,  {Meru} F.,  {Lee} M.~H.,   {Clarke} C.~J.,  2018, \mn@doi
  [\mnras] {10.1093/mnras/stx3084}, \href
  {http://adsabs.harvard.edu/abs/2018MNRAS.475..167B} {475, 167}

\bibitem[\protect\citeauthoryear{{Chiang} \& {Goldreich}}{{Chiang} \&
  {Goldreich}}{1997}]{CG97}
{Chiang} E.~I.,  {Goldreich} P.,  1997, \mn@doi [\apj] {10.1086/304869}, \href
  {http://adsabs.harvard.edu/abs/1997ApJ...490..368C} {490, 368}

\bibitem[\protect\citeauthoryear{{Chiang} \& {Laughlin}}{{Chiang} \&
  {Laughlin}}{2013}]{chiang13_mmsn}
{Chiang} E.,  {Laughlin} G.,  2013, \mn@doi [\mnras] {10.1093/mnras/stt424},
  \href {http://adsabs.harvard.edu/abs/2013MNRAS.431.3444C} {431, 3444}

\bibitem[\protect\citeauthoryear{{Chiang} \& {Youdin}}{{Chiang} \&
  {Youdin}}{2010}]{CY10_review}
{Chiang} E.,  {Youdin} A.~N.,  2010, \mn@doi [Annual Review of Earth and
  Planetary Sciences] {10.1146/annurev-earth-040809-152513}, \href
  {http://adsabs.harvard.edu/abs/2010AREPS..38..493C} {38, 493}

\bibitem[\protect\citeauthoryear{{{\'C}uk} \& {Burns}}{{{\'C}uk} \&
  {Burns}}{2004}]{Cuk04}
{{\'C}uk} M.,  {Burns} J.~A.,  2004, \mn@doi [\icarus]
  {10.1016/j.icarus.2003.09.026}, \href
  {http://adsabs.harvard.edu/abs/2004Icar..167..369C} {167, 369}

\bibitem[\protect\citeauthoryear{{De Simone}, {Wu}  \& {Tremaine}}{{De Simone}
  et~al.}{2004}]{tre04}
{De Simone} R.,  {Wu} X.,   {Tremaine} S.,  2004, \mn@doi [\mnras]
  {10.1111/j.1365-2966.2004.07675.x}, \href
  {http://adsabs.harvard.edu/abs/2004MNRAS.350..627D} {350, 627}

\bibitem[\protect\citeauthoryear{{Dermott}, {Christou}, {Li}, {Kehoe}  \&
  {Robinson}}{{Dermott} et~al.}{2018}]{Der+18}
{Dermott} S.~F.,  {Christou} A.~A.,  {Li} D.,  {Kehoe} T.~J.~J.,   {Robinson}
  J.~M.,  2018, \mn@doi [Nature Astronomy] {10.1038/s41550-018-0482-4}, \href
  {http://adsabs.harvard.edu/abs/2018NatAs...2..549D} {2, 549}

\bibitem[\protect\citeauthoryear{{Do}, {Tucker}  \& {Tonry}}{{Do}
  et~al.}{2018}]{Do18}
{Do} A.,  {Tucker} M.~A.,   {Tonry} J.,  2018, preprint, \href
  {http://adsabs.harvard.edu/abs/2018arXiv180102821D} {} (\mn@eprint {arXiv}
  {1801.02821})

\bibitem[\protect\citeauthoryear{{Dohnanyi}}{{Dohnanyi}}{1969}]{d69}
{Dohnanyi} J.~S.,  1969, \mn@doi [\jgr] {10.1029/JB074i010p02531}, \href
  {http://adsabs.harvard.edu/abs/1969JGR....74.2531D} {74, 2531}

\bibitem[\protect\citeauthoryear{{Dones}, {Gladman}, {Melosh}, {Tonks},
  {Levison}  \& {Duncan}}{{Dones} et~al.}{1999}]{Dones99}
{Dones} L.,  {Gladman} B.,  {Melosh} H.~J.,  {Tonks} W.~B.,  {Levison} H.~F.,
  {Duncan} M.,  1999, \mn@doi [\icarus] {10.1006/icar.1999.6220}, \href
  {http://adsabs.harvard.edu/abs/1999Icar..142..509D} {142, 509}

\bibitem[\protect\citeauthoryear{{Fujita}, {Ohtsuki}, {Tanigawa}  \&
  {Suetsugu}}{{Fujita} et~al.}{2013}]{Fujita13}
{Fujita} T.,  {Ohtsuki} K.,  {Tanigawa} T.,   {Suetsugu} R.,  2013, \mn@doi
  [\aj] {10.1088/0004-6256/146/6/140}, \href
  {http://adsabs.harvard.edu/abs/2013AJ....146..140F} {146, 140}

\bibitem[\protect\citeauthoryear{{Grishin} \& {Perets}}{{Grishin} \&
  {Perets}}{2015}]{GP15}
{Grishin} E.,  {Perets} H.~B.,  2015, \mn@doi [\apj]
  {10.1088/0004-637X/811/1/54}, \href
  {http://adsabs.harvard.edu/abs/2015ApJ...811...54G} {811, 54}

\bibitem[\protect\citeauthoryear{{Grishin} \& {Perets}}{{Grishin} \&
  {Perets}}{2016}]{GP16}
{Grishin} E.,  {Perets} H.~B.,  2016, \mn@doi [\apj]
  {10.3847/0004-637X/820/2/106}, \href
  {http://adsabs.harvard.edu/abs/2016ApJ...820..106G} {820, 106}

\bibitem[\protect\citeauthoryear{{Hayashi}}{{Hayashi}}{1981}]{Hayashi81}
{Hayashi} C.,  1981, \mn@doi [Progress of Theoretical Physics Supplement]
  {10.1143/PTPS.70.35}, \href
  {http://adsabs.harvard.edu/abs/1981PThPS..70...35H} {70, 35}

\bibitem[\protect\citeauthoryear{{Hut}, {Makino}  \& {McMillan}}{{Hut}
  et~al.}{1995}]{Hut95}
{Hut} P.,  {Makino} J.,   {McMillan} S.,  1995, \mn@doi [\apjl]
  {10.1086/187844}, \href {http://adsabs.harvard.edu/abs/1995ApJ...443L..93H}
  {443, L93}

\bibitem[\protect\citeauthoryear{{Johansen}, {Oishi}, {Mac Low}, {Klahr},
  {Henning}  \& {Youdin}}{{Johansen} et~al.}{2007}]{J07}
{Johansen} A.,  {Oishi} J.~S.,  {Mac Low} M.-M.,  {Klahr} H.,  {Henning} T.,
  {Youdin} A.,  2007, \mn@doi [\nat] {10.1038/nature06086}, \href
  {http://adsabs.harvard.edu/abs/2007Natur.448.1022J} {448, 1022}

\bibitem[\protect\citeauthoryear{{Lambrechts} \& {Johansen}}{{Lambrechts} \&
  {Johansen}}{2012}]{LJ12}
{Lambrechts} M.,  {Johansen} A.,  2012, \mn@doi [\aap]
  {10.1051/0004-6361/201219127}, \href
  {http://adsabs.harvard.edu/abs/2012A26A...544A..32L} {544, A32}

\bibitem[\protect\citeauthoryear{{Levison}, {Duncan}, {Brasser}  \&
  {Kaufmann}}{{Levison} et~al.}{2010}]{Lev+10}
{Levison} H.~F.,  {Duncan} M.~J.,  {Brasser} R.,   {Kaufmann} D.~E.,  2010,
  \mn@doi [Science] {10.1126/science.1187535}, \href
  {http://adsabs.harvard.edu/abs/2010Sci...329..187L} {329, 187}

\bibitem[\protect\citeauthoryear{{Levison}, {Kretke}  \& {Duncan}}{{Levison}
  et~al.}{2015}]{Levison15}
{Levison} H.~F.,  {Kretke} K.~A.,   {Duncan} M.~J.,  2015, \mn@doi [\nat]
  {10.1038/nature14675}, \href
  {http://adsabs.harvard.edu/abs/2015Natur.524..322L} {524, 322}

\bibitem[\protect\citeauthoryear{{Lingam} \& {Loeb}}{{Lingam} \&
  {Loeb}}{2018}]{Loeb18}
{Lingam} M.,  {Loeb} A.,  2018, preprint, \href
  {http://adsabs.harvard.edu/abs/2018arXiv180110254L} {} (\mn@eprint {arXiv}
  {1801.10254})

\bibitem[\protect\citeauthoryear{{Meech} et~al.,}{{Meech}
  et~al.}{2017}]{Meech17}
{Meech} K.~J.,  et~al., 2017, \mn@doi [\nat] {10.1038/nature25020}, \href
  {http://adsabs.harvard.edu/abs/2017Natur.552..378M} {552, 378}

\bibitem[\protect\citeauthoryear{{Melosh}}{{Melosh}}{2003}]{Melosh2003}
{Melosh} H.~J.,  2003, \mn@doi [Astrobiology] {10.1089/153110703321632525},
  \href {http://adsabs.harvard.edu/abs/2003AsBio...3..207M} {3, 207}

\bibitem[\protect\citeauthoryear{{Morbidelli}}{{Morbidelli}}{2018}]{Mor18}
{Morbidelli} A.,  2018, ArXiv:1803.06704, \href
  {http://adsabs.harvard.edu/abs/2018arXiv180306704M} {}

\bibitem[\protect\citeauthoryear{{Morbidelli}, {Bottke}, {Nesvorn{\'y}}  \&
  {Levison}}{{Morbidelli} et~al.}{2009}]{Mor+09}
{Morbidelli} A.,  {Bottke} W.~F.,  {Nesvorn{\'y}} D.,   {Levison} H.~F.,  2009,
  \mn@doi [\icarus] {10.1016/j.icarus.2009.07.011}, \href
  {http://adsabs.harvard.edu/abs/2009Icar..204..558M} {204, 558}

\bibitem[\protect\citeauthoryear{{Napier}}{{Napier}}{2004}]{Napier}
{Napier} W.~M.,  2004, \mn@doi [\mnras] {10.1111/j.1365-2966.2004.07287.x},
  \href {http://adsabs.harvard.edu/abs/2004MNRAS.348...46N} {348, 46}

\bibitem[\protect\citeauthoryear{{Ormel} \& {Klahr}}{{Ormel} \&
  {Klahr}}{2010}]{Orm+10}
{Ormel} C.~W.,  {Klahr} H.~H.,  2010, \mn@doi [\aap]
  {10.1051/0004-6361/201014903}, \href
  {http://adsabs.harvard.edu/abs/2010A%26A...520A..43O} {520, A43}

\bibitem[\protect\citeauthoryear{{Ormel} \& {Kobayashi}}{{Ormel} \&
  {Kobayashi}}{2012}]{Ormel+12}
{Ormel} C.~W.,  {Kobayashi} H.,  2012, \mn@doi [\apj]
  {10.1088/0004-637X/747/2/115}, \href
  {http://adsabs.harvard.edu/abs/2012ApJ...747..115O} {747, 115}

\bibitem[\protect\citeauthoryear{{Ouellette}, {Desch}  \& {Hester}}{{Ouellette}
  et~al.}{2010}]{Ouellette2010}
{Ouellette} N.,  {Desch} S.~J.,   {Hester} J.~J.,  2010, \mn@doi [\apj]
  {10.1088/0004-637X/711/2/597}, \href
  {http://adsabs.harvard.edu/abs/2010ApJ...711..597O} {711, 597}

\bibitem[\protect\citeauthoryear{{Perets} \& {Kouwenhoven}}{{Perets} \&
  {Kouwenhoven}}{2012}]{Perets+12}
{Perets} H.~B.,  {Kouwenhoven} M.~B.~N.,  2012, \mn@doi [\apj]
  {10.1088/0004-637X/750/1/83}, \href
  {http://adsabs.harvard.edu/abs/2012ApJ...750...83P} {750, 83}

\bibitem[\protect\citeauthoryear{{Perets} \& {Murray-Clay}}{{Perets} \&
  {Murray-Clay}}{2011}]{PMC11}
{Perets} H.~B.,  {Murray-Clay} R.~A.,  2011, \mn@doi [\apj]
  {10.1088/0004-637X/733/1/56}, \href
  {http://adsabs.harvard.edu/abs/2011ApJ...733...56P} {733, 56}

\bibitem[\protect\citeauthoryear{{Pinhas}, {Madhusudhan}  \& {Clarke}}{{Pinhas}
  et~al.}{2016}]{Ar16}
{Pinhas} A.,  {Madhusudhan} N.,   {Clarke} C.,  2016, \mn@doi [\mnras]
  {10.1093/mnras/stw2239}, \href
  {http://adsabs.harvard.edu/abs/2016MNRAS.463.4516P} {463, 4516}

\bibitem[\protect\citeauthoryear{{Raymond} \& {Cossou}}{{Raymond} \&
  {Cossou}}{2014}]{raymond14_mmsn}
{Raymond} S.~N.,  {Cossou} C.,  2014, \mn@doi [\mnras] {10.1093/mnrasl/slu011},
  \href {http://adsabs.harvard.edu/abs/2014MNRAS.440L..11R} {440, L11}

\bibitem[\protect\citeauthoryear{{Raymond}, {Armitage}, {Veras}, {Quintana}  \&
  {Barclay}}{{Raymond} et~al.}{2018a}]{Raymond17}
{Raymond} S.~N.,  {Armitage} P.~J.,  {Veras} D.,  {Quintana} E.~V.,   {Barclay}
  T.,  2018a, \mn@doi [\mnras] {10.1093/mnras/sty468}, \href
  {http://adsabs.harvard.edu/abs/2018MNRAS.476.3031R} {476, 3031}

\bibitem[\protect\citeauthoryear{{Raymond}, {Armitage}  \& {Veras}}{{Raymond}
  et~al.}{2018b}]{Raymond18}
{Raymond} S.~N.,  {Armitage} P.~J.,   {Veras} D.,  2018b, \mn@doi [\apjl]
  {10.3847/2041-8213/aab4f6}, \href
  {http://adsabs.harvard.edu/abs/2018ApJ...856L...7R} {856, L7}

\bibitem[\protect\citeauthoryear{{Simon}, {Armitage}, {Youdin}  \&
  {Li}}{{Simon} et~al.}{2017}]{Simon17}
{Simon} J.~B.,  {Armitage} P.~J.,  {Youdin} A.~N.,   {Li} R.,  2017, \mn@doi
  [\apjl] {10.3847/2041-8213/aa8c79}, \href
  {http://adsabs.harvard.edu/abs/2017ApJ...847L..12S} {847, L12}

\bibitem[\protect\citeauthoryear{{Valtonen} et~al.,}{{Valtonen}
  et~al.}{2009}]{Valtonen09}
{Valtonen} M.,  et~al., 2009, \mn@doi [\apj] {10.1088/0004-637X/690/1/210},
  \href {http://adsabs.harvard.edu/abs/2009ApJ...690..210V} {690, 210}

\bibitem[\protect\citeauthoryear{{Weidenschilling}}{{Weidenschilling}}{1977a}]{wei77_mmsn}
{Weidenschilling} S.~J.,  1977a, \mn@doi [A\&A, Supplement]
  {10.1007/BF00642464}, \href
  {http://adsabs.harvard.edu/abs/1977Ap%26SS..51..153W} {51, 153}

\bibitem[\protect\citeauthoryear{{Weidenschilling}}{{Weidenschilling}}{1977b}]{Wei77}
{Weidenschilling} S.~J.,  1977b, \mn@doi [\mnras] {10.1093/mnras/180.1.57},
  \href {http://adsabs.harvard.edu/abs/1977MNRAS.180...57W} {180, 57}

\bibitem[\protect\citeauthoryear{{Wesson}}{{Wesson}}{2010}]{Wes+10}
{Wesson} P.~S.,  2010, \mn@doi [\ssr] {10.1007/s11214-010-9671-x}, \href
  {http://adsabs.harvard.edu/abs/2010SSRv..156..239W} {156, 239}

\bibitem[\protect\citeauthoryear{{Windmark}, {Birnstiel}, {G{\"u}ttler},
  {Blum}, {Dullemond}  \& {Henning}}{{Windmark} et~al.}{2012}]{Win+12a}
{Windmark} F.,  {Birnstiel} T.,  {G{\"u}ttler} C.,  {Blum} J.,  {Dullemond}
  C.~P.,   {Henning} T.,  2012, \mn@doi [\aap] {10.1051/0004-6361/201118475},
  \href {http://adsabs.harvard.edu/abs/2012A%26A...540A..73W} {540, A73}

\bibitem[\protect\citeauthoryear{{Xie}, {Payne}, {Th{\'e}bault}, {Zhou}  \&
  {Ge}}{{Xie} et~al.}{2010}]{Xie+18}
{Xie} J.-W.,  {Payne} M.~J.,  {Th{\'e}bault} P.,  {Zhou} J.-L.,   {Ge} J.,
  2010, \mn@doi [\apj] {10.1088/0004-637X/724/2/1153}, \href
  {http://adsabs.harvard.edu/abs/2010ApJ...724.1153X} {724, 1153}

\bibitem[\protect\citeauthoryear{{Youdin} \& {Goodman}}{{Youdin} \&
  {Goodman}}{2005}]{Y05}
{Youdin} A.~N.,  {Goodman} J.,  2005, \mn@doi [\apj] {10.1086/426895}, \href
  {http://adsabs.harvard.edu/abs/2005ApJ...620..459Y} {620, 459}

\makeatother
\end{thebibliography}

\bibliographystyle{mnras}

\section*{Acknowledgments}
We thank Barak A. Katzir and Andrei P. Igoshev for stimulating discussions. EG acknowledges
support by the Technion Irwin and Joan Jacobs Excellence Fellowship for outstanding graduate
students. EG and HBP acknowledge support by Israel Science Foundation I-CORE grant
1829/12 and the Minerva center for life under exreme planetary conditions.

\subsection*{Appendix A. Derivation of the Gravitational focusing regime}

Here we Derive Eq. \ref{eq:fx} and \ref{eq:cdf2}. We start from a $\chi(4)$ distribution for the velocity:
\begin{equation}
\tilde{f}_{V}(v)=\frac{1}{2}\frac{v^{3}}{\sigma^{4}}e^{-v^{2}/2\sigma^{2}}
\end{equation}
and uniform distribution of $b^2$, namely $f_B(b) \propto b$. 
In order to proceed, we define a new random variable $x\equiv(b/b_{\rm max})^{\alpha}(v_{\infty}/\sigma)^{2}$,
$f_{X}(x)$, with $\alpha$ is related to the disc power law density and defined in the main text.

First, we transform to $k=b^{\alpha}$ and ${\rm d}k=\alpha b^{\alpha-1}{\rm d}b$,
so 
\begin{equation}
f_{K}(k)=f_{B}(b)\frac{{\rm d}b}{{\rm d}k}=\frac{2k^{(2-\alpha)/\alpha}}{\alpha b_{\rm max}^{2}}.
\end{equation}
Similarly, the distribution function of $u\equiv v^{2}$ is 
\begin{equation}
\tilde{f}_{U}(u)=\tilde{f}_{V}(v)\frac{{\rm d}v}{{\rm d}u}=\frac{1}{4}\frac{u}{\sigma^{4}}e^{-u/2\sigma^{2}}
\end{equation}
Now, the distribution of $z\equiv b^{\alpha}v^{2}$ is given by
\begin{align}
f_{Z}(z) & =  \iint \tilde{f}_{U}(t)f_{K}(k)\delta(ku-z){\rm d}k{\rm d}u \nonumber \\
 & =  \frac{1}{2 \alpha \sigma^{4} b_{\rm max}^{2}} \iint k^{(2-\alpha)/\alpha} e^{-u/2\sigma^{2}}\delta(k-z/u){\rm d}k{\rm d}u \nonumber \\
 & = \frac{z^{(2-\alpha)/\alpha}}{2\alpha\sigma^{4}b_{{\rm max}}^{2}}\intop_{z/k_{{\rm max}}}^{\infty}u^{(\alpha-2)/\alpha}e^{-u/2\sigma^{2}}{\rm d}u, \label{eq:fz}
\end{align}
where $\delta(s)$ is Dirac's delta distribution. Taking $x=z/b_{{\rm max}}^{\alpha}\sigma^{2}$ and $w=u/2\sigma^{2}$
we get 
\begin{align}
f_{X}(x) & =\frac{x^{(2-\alpha)/\alpha}}{\alpha}2^{1-2/\alpha}\intop_{x/2}^{\infty}w^{1-2/\alpha}e^{-w}{\rm d}w \nonumber \\
 & =2^{-(2+\beta)/(1+\beta)}\frac{2+\beta}{1+\beta}\Gamma\left(\frac{\beta}{1+\beta},\frac{x}{2}\right)x^{1/(1+\beta)}
\end{align}
where $\Gamma(s,z) = \int_{z}^{\infty}t^{s-1}e^{-t}{\rm d}t$ is the upper incomplete Gamma function. By using the integrals of the incomplete Gamma functions
\begin{equation}
\int x^{b-1}\Gamma(s,x){\rm d}x = \frac{1}{b}\left( x^b\Gamma(s,x) - \Gamma(s+b,x) \right),
\end{equation}
the cumulative distribution function (CDF) is given by 
\begin{align}
F_X(x) = \int f_X(x) {\rm d}x  = \left(\frac{x}{2}\right)^{\frac{2+\beta}{1+\beta}}\Gamma\left(\frac{\beta}{1+\beta},\frac{x}{2}\right) - \Gamma\left(2,\frac{x}{2}\right).
\end{align}
The linear expansion of the incomplete Gamma function is $\Gamma\left(s,\frac{x}{2}  \right) = \Gamma(s) - \frac{x^s}{2^{-s}s} + \pazocal{O}\left( x^{s+1} \right)$, therefore the leading term in CDF is 
\begin{equation}
F_X(x) = 2^{-\frac{2+\beta}{1+\beta}}x^{\frac{2+\beta}{1+\beta}} \Gamma\left(\frac{\beta}{1+\beta}\right) -\frac{1+\beta}{\beta}2^{-\frac{2}{1+\beta}}x^{2} - 1 + 2x^2.
\end{equation}
For $\beta>0$, the first term is the dominant one, thus the capture probability is 
\begin{equation}
f_c(R_p) = F_X(x) - F_X(0) = 2^{-\frac{2+\beta}{1+\beta}}x^{\frac{2+\beta}{1+\beta}} \Gamma\left(\frac{\beta}{1+\beta}\right) + \pazocal{O}\left(x^2\right)
\end{equation}

\subsection*{Appendix B. Planetesimal ablation}

\begin{figure}
\begin{centering}
\includegraphics[width=8.5cm]{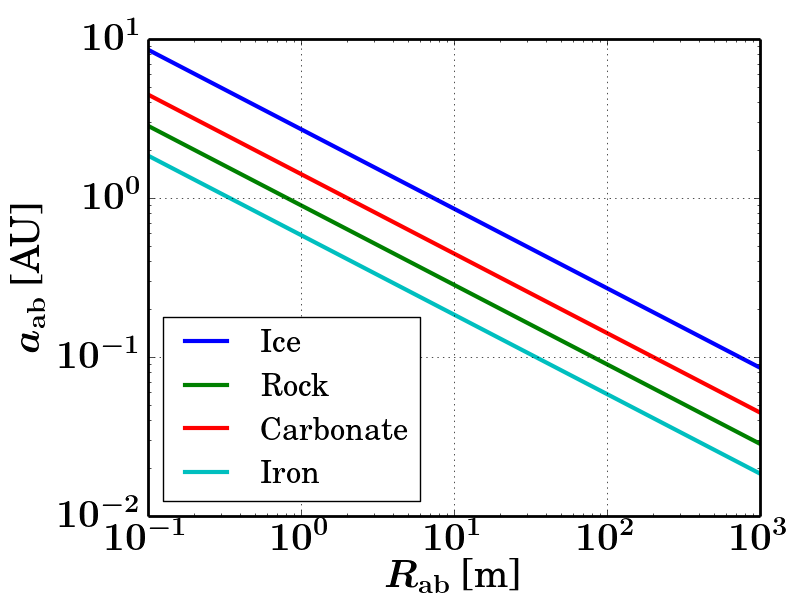}
\caption{\label{fig:A1} Critical separations for significant planetesimal ablation. Specific ablation heat coefficients and densities are taken from Table 1 of \protect\cite{Ar16}  }
\end{centering}
\end{figure}

The ablation equation is \citep{Ar16}
\begin{equation}
\frac{dm}{dt}=-\frac{C_{H}}{2}\frac{\rho_{g}v_{\rm rel}^{3}\pi R_p^{2}}{Q_{\rm abl}}
\end{equation}
where $m=4\pi R_p^{3}\rho_{p}/3$ is the mass, $\rho_{p}$ is the solid density, $\rho_{g}$ is the gas density, $C_{H}$ is the dimensionless heat transfer coefficient, $v_{\rm rel}$ is the relative velocity, $R_p$ is the radius and $Q_{\rm abl}$ is the specific ablation heat per unit mass. The ablation time is 
\begin{equation}
t_{\rm abl}=\left|\frac{R_p}{dR_p/dt}\right|\approx\frac{8}{C_{H}}\frac{\rho_{p}}{\rho_{g}}\frac{Q_{\rm abl}}{v_{\rm rel}^{3}}R_p \approx 10^{4}\left(\frac{R_p}{m}\right)\left(\frac{a}{\rm AU}\right)^{53/14}\ \rm s
\end{equation}
Significant ablation occurs if the ablation times is shorter than the minimum of the disc crossing time $t_{\rm cross}$ and the stopping time $t_{\rm stop} = | m v_{\rm rel} / F_{\rm D}|$. For pebbles of $\gtrsim 1\ \rm m$ or larger bodies, the crossing time $t_{\rm cross}=h/v_{\rm rel}$ is the relevant. Comparing the timescale gives the condition for ablation. The critical radial disc separation for ablation as a function of the planetesimal size and disc and planetesimal parameters is
\begin{equation}
a_{\rm ab}=\sqrt{\frac{C_{H}}{8}\frac{\rho_{g}}{\rho_{p}}\frac{h}{R_p}\frac{v_{\rm rel}^2}{Q_{\rm abl}}}\ \rm AU
\end{equation} 
For typical compositions of ices (see Table 1 \citealp{Ar16}), $C_H=0.01$, $Q_{\rm abl}\approx 3 \cdot 10^{10}\ \rm erg\ g^{-1}$, $\rho_p = 1\ \rm g\ cm^{-3}$ and $\rho_p$ and $v_{\rm rel}\approx v_{\rm esc}$ normalized to their values at $1\ \rm AU$ the critical radial separation is 

\begin{equation}
a_{\rm ab}=\left(\frac{R}{7.3\ \rm m}\right)^{-1/2}\ \rm AU
\end{equation} 

Figure \ref{fig:A1} shows the critical separation as a function of the planetesimal size for various compositions. We compare the critical separation for each composition with the cumulative fraction of captured planetesimal to estimate the fraction of ablated planetesimals. The results are summarized in table \ref{tab1}.


  
\end{document}